\documentclass[]{spie}  

\usepackage{amsmath,amsfonts,amssymb}
\usepackage{graphicx}
\usepackage[colorlinks=true, allcolors=blue]{hyperref}
\usepackage{amssymb}
\usepackage{adjustbox}
\usepackage{graphicx}
\usepackage{siunitx}
\usepackage{mathtools}
\usepackage{threeparttable}
\usepackage{float}
\usepackage{booktabs}
\usepackage{tabularx}
\usepackage{makecell}
\newcolumntype{C}[1]{>{\centering\arraybackslash}p{#1}}
\newcolumntype{L}[1]{>{\raggedright\arraybackslash}p{#1}}
\usepackage{multicol}
\usepackage[T1]{fontenc}
\usepackage{csquotes}
\usepackage{url}
\usepackage{subcaption}

\title{Lymph Node Detection in T2 MRI with Transformers}

\author[a]{Tejas Sudharshan Mathai}
\author[a]{Sungwon Lee}
\author[a]{Daniel C. Elton}
\author[a]{Thomas C. Shen}
\author[b]{Yifan Peng}
\author[b]{Zhiyong Lu}
\author[a]{Ronald M. Summers}
\affil[a]{Imaging Biomarkers and Computer-Aided Diagnosis Laboratory, Radiology and Imaging Sciences, Clinical Center, National Institutes of Health, Bethesda MD, USA}
\affil[b]{National Center for Biotechnology Information, National Library of Medicine, National Institutes of Health, Bethesda, MD, USA}

\authorinfo{Further author information: (Send correspondence to T.S.M.)\\T.S.M.: E-mail: tejas.mathai@nih.gov}

\graphicspath{ 
{fig_money/} 
{fig_results/}
{fig_results/12899_6_006/}
{fig_results/09173_5_021/}
{fig_results/11933_5_015/}
{fig_results/19028_5_006/}
}

\begin{document} 

\maketitle

\begin{abstract}
Identification of lymph nodes (LN) in T2 Magnetic Resonance Imaging (MRI) is an important step performed by radiologists during the assessment of lymphoproliferative diseases. The size of the nodes play a crucial role in their staging, and radiologists sometimes use an additional contrast sequence such as diffusion weighted imaging (DWI) for confirmation. However, lymph nodes have diverse appearances in T2 MRI scans, making it tough to stage for metastasis. Furthermore, radiologists often miss smaller metastatic lymph nodes over the course of a busy day. To deal with these issues, we propose to use the DEtection TRansformer (DETR) network to localize suspicious metastatic lymph nodes for staging in challenging T2 MRI scans acquired by different scanners and exam protocols. False positives (FP) were reduced through a bounding box fusion technique, and a precision of 65.41\% and sensitivity of 91.66\% at 4 FP per image was achieved. To the best of our knowledge, our results improve upon the current state-of-the-art for lymph node detection in T2 MRI scans.
\end{abstract}




\section{INTRODUCTION}
\label{intro}  

Lymph nodes (LN) are a part of the lymphatic system that helps the body fight infection by removing foreign substances. Enlarged and metastatic LN require accurate identification especially when they are present at sites that do not correspond to the first site of lymphatic spread, as it indicates distant metastasis \cite{Amin2017}. These enlarged LN can be sized according to the AJCC guidelines \cite{Amin2017}, which cover the management of cancer and lymphoproliferative diseases. LN are usually imaged with multi-parametric MRI, e.g. T2 and Diffusion Weighted Imaging (DWI), but the type of scanners used and exam protocols preferred span the gamut across different institutions. Moreover, their diverse appearance, irregular shapes, and varying anatomical location make it difficult to stage them. Nodal size is usually measured with two orthogonal lines: the long and short axis diameters (LAD and SAD), and both are usually required for charting the course of therapy; however, these guidelines also vary by institution. Due to these imaging and workflow issues, there is a need for automated LN detection in T2 MRI images for sizing.

There is limited prior on the LN detection task in MRI scans \cite{Zhao2020_mri,Debats2019_mri} whereas the majority of LN detection research has been focused towards CT scans \cite{Liu2016_ct,Seff2015_ct,Roth2014_ct}. Typically, T2 MRI and DWI are preferred in clinical practice for LN staging, but for algorithmic development, registration of DWI to T2 scans is needed \cite{Zhao2020_mri}. In this work, we focus only on identifying LN in challenging T2 MRI scans acquired with different scanners and exam protocols, quantify the detection performance of state-of-the-art one-stage detection networks \cite{Tian2019_fcos,Kong2019_foveabox,Zhang2021_vfnet}, and surpass them with the recently published DEtection TRansformer (DETR) network. Our intent is centered on the belief that the reliable detection of LN in T2 MRI scans can be supplemented by DWI scans later. We also boost the detection performance with bounding box fusion techniques \cite{Solovyev2021} to decrease the false positive rate. Our results exceed the capabilities of the previous lymph node detection approaches with a precision of 65.41\% and sensitivity of 91.66\% at 4 FP per image.

\section{METHODS}
\label{methods}  


\noindent
\textbf{Anchor-Free One-Stage Detectors.} We quantified the performance of state-of-the-art one-stage anchor-free object detectors on the LN detection task in T2 MRI: 1) FCOS \cite{Tian2019_fcos}, 2) FoveaBox \cite{Kong2019_foveabox}, and 3) VFNet \cite{Zhang2021_vfnet}. These detectors are superior to anchor-based detectors (e.g. RetinaNet) and two-stage detectors (e.g Faster RCNN) because they skip the region proposal stage, and directly predict the bounding box coordinates and class probabilities for different categories in a single pass. FCOS \cite{Tian2019_fcos} employs multi-level prediction of feature maps for object detection inside a Fully Convolutional Network (FCN). It also computes a centerness score to reduce the FP, which tend to be far away from the target object center. VFNet combines FCOS (without centerness branch) with efficient sample selection, integrates a classification score based on the IoU value between the ground truth and the prediction into a novel IoU-aware Varifocal loss, and refines the bounding box predictions. FoveaBox \cite{Kong2019_foveabox} has a ResNet50 backbone to compute input image features that is used by a fovea head network, which estimates the object occurrence possibility through per-pixel classification on the backbone's output. 

\noindent
\textbf{DEtection Transformer (DETR).} DETR \cite{Carion2020_detr} uses the bipartite set matching loss and parallel decoding to detect LN. It uses a ResNet50 backbone to compute image features and adds spatial positional encoding to them, feeds the features into an encoder-decoder architecture with self-attention modules, and finally uses feed-forward network heads that uniquely assign predictions to the ground truth in parallel. Different from the original implementation \cite{Carion2020_detr}, we replaced cross entropy loss as the classification loss in the bounding box matching cost with the focal loss \cite{Lin2017_retinanet} to overcome the class imbalance problem and weighted it with $\lambda_{c}=2$.

\begin{figure}[!hb]
\centering
\begin{subfigure}[b]{0.19\columnwidth}
\vspace*{\fill}
  \centering
  \includegraphics[width=\columnwidth,height=2.5cm]{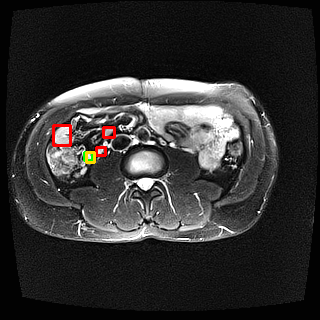}
  \includegraphics[width=\columnwidth,height=2.5cm]{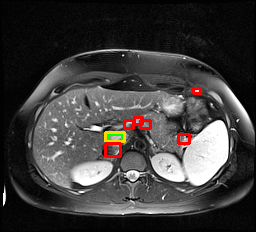}
  \includegraphics[width=\columnwidth,height=2.5cm]{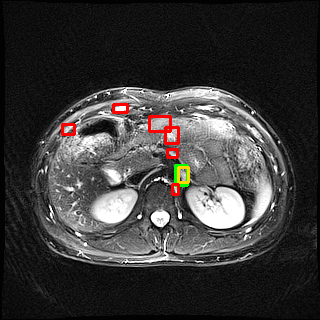}
  \includegraphics[width=\columnwidth,height=2.5cm]{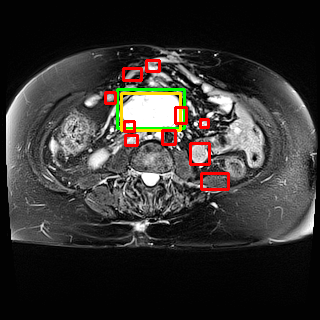}
  \centerline{(a) FCOS}
\end{subfigure} 
\begin{subfigure}[b]{0.19\columnwidth}
\vspace*{\fill}
  \centering
  \includegraphics[width=\columnwidth,height=2.5cm]{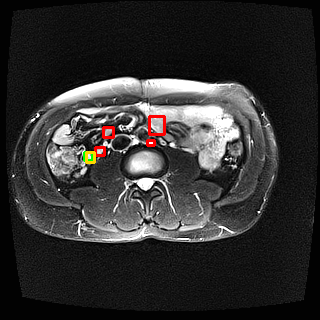}
  \includegraphics[width=\columnwidth,height=2.5cm]{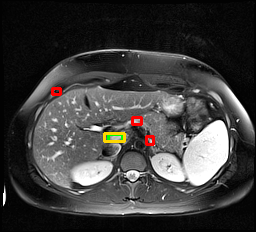}
  \includegraphics[width=\columnwidth,height=2.5cm]{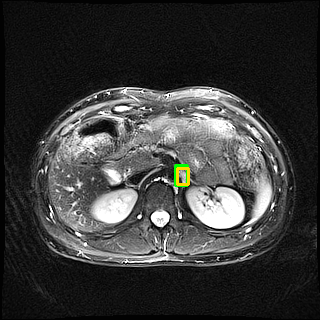}
  \includegraphics[width=\columnwidth,height=2.5cm]{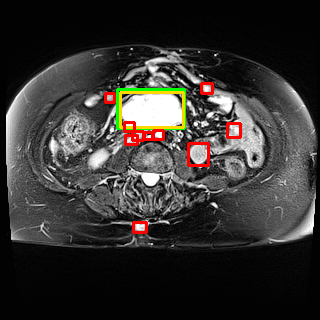}
  \centerline{(b) FoveaBox}
\end{subfigure} 
\begin{subfigure}[b]{0.19\columnwidth}
\vspace*{\fill}
  \centering
  \includegraphics[width=\columnwidth,height=2.5cm]{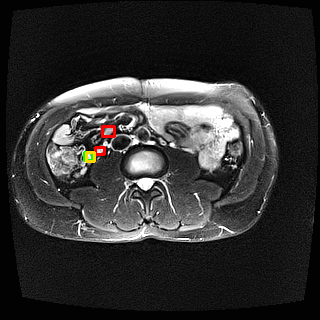}
  \includegraphics[width=\columnwidth,height=2.5cm]{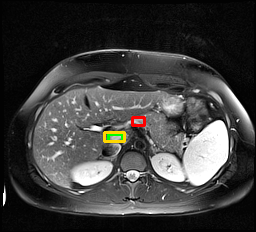}
  \includegraphics[width=\columnwidth,height=2.5cm]{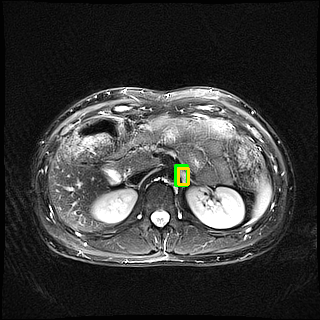}
  \includegraphics[width=\columnwidth,height=2.5cm]{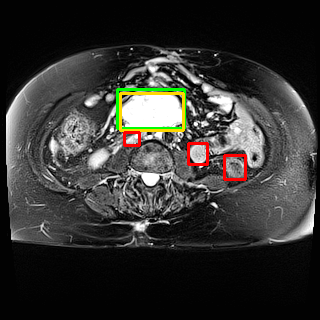}
  \centerline{(c) VFNet}
\end{subfigure} 
\begin{subfigure}[b]{0.19\columnwidth}
\vspace*{\fill}
  \centering
  \includegraphics[width=\columnwidth,height=2.5cm]{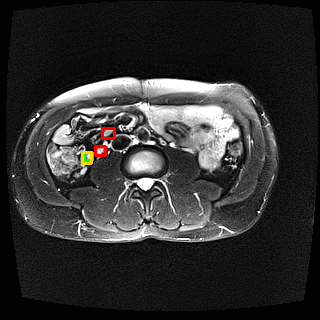}
  \includegraphics[width=\columnwidth,height=2.5cm]{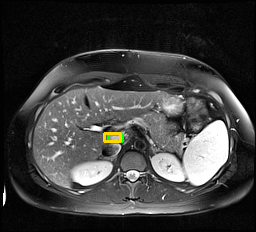}
  \includegraphics[width=\columnwidth,height=2.5cm]{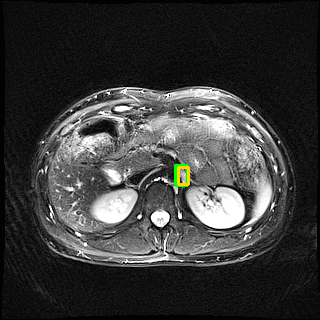}
  \includegraphics[width=\columnwidth,height=2.5cm]{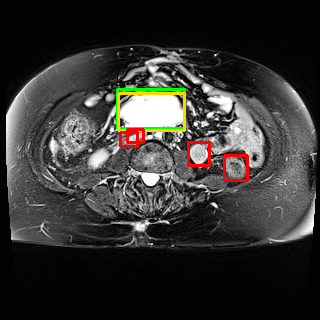}
  \centerline{(d) Proposed DETR}
\end{subfigure} 
\caption{Columns (a) - (c) show the lymph node detection results of the one-stage detectors (FCOS, FoveaBox, VFNet) on four different T2 MRI images. Column (d) displays the result of our proposed DETR transformer following Weighted Boxes Fusion. Green boxes: ground truth, yellow: true positives, and red: false positives.}
\label{fig:qual_images}
\end{figure}

\noindent
\textbf{Weighted Boxes Fusion.} Five epochs from each detector with the lowest validation loss were used for prediction, thereby generating multiple bounding box predictions (with confidence scores) at the nodal location. Some of these predictions clustered together in common regions in the image, and Weighted Boxes Fusion \cite{Solovyev2021} was employed to combine the clusters and yield precise predictions (as seen in Fig. \ref{fig:qual_images}). 

\section{EXPERIMENTS AND RESULTS}
\label{results}

\subsection{Data}

The lymph node dataset contained abdominal MRI scans downloaded from the National Institutes of Health (NIH) Picture Archiving and Communication System (PACS), and were acquired between January 2015 and September 2019. Initially, 584 T2-weighted MRI scans and associated radiology reports from different patients ($n$=584) were downloaded. The nodal extent and size measurements were extracted from the reports using NLP \cite{Peng2020}. An experienced radiologist checked the collected data, and removed incorrect annotations and scans containing only one LN annotation (either LAD or SAD measures). This resulted in a total of 376 T2 scans with 520 distinct LN that had both the LAD and SAD measures. The voxels in the scans were normalized to the [1\%, 99\%] of their intensity range \cite{Kociolek2020} so as to boost contrast between bright and dark structures, and finally histogram equalized. The final dataset was then randomly divided into training (60\%, 225 scans), validation (20\%, 76 scans), and test (20\%, 75 scans) splits at the patient-level. The resulting scans had dimensions in the range from (256 $\sim$ 640) $\times$ (192 $\sim$ 640) $\times$ (18 $\sim$ 60) voxels.

\subsection{Implementation}

When reading scans, radiologists scroll through 1-3 slices of the volume, determine the extent of the lymph nodes, and verify the finding using another sequence (e.g. DWI). To mimic their workflow, we use 3-slice T2 MRI images with the center slice containing the radiologist annotated LN as the input to a detection network framework \cite{Chen2019_mmdet}. The backbone for the one-stage detectors (FCOS, FoveaBox, VFNet) was ResNet50 (pre-trained with MS COCO weights) and standard data augmentation was performed: random flips, crops, shifts and rotations in the range of [0, 32] pixels, and [0, 10] degrees respectively. To train the one-stage detectors, we used a batch size of 2, learning rate of 1e-3, and trained each model for 24 epochs. The 5 epochs with the lowest validation loss were ensembled together and used for LN detection. For DETR, we used a learning rate of 1.25e-05 and trained the model for 150 epochs. All experiments were run on a workstation running Ubuntu 16.04LTS and containing a NVIDIA Tesla V100 GPU. Our results are shown in Table \ref{table_LN_detection_results}.

\subsection{Results}

Among the various one-stage detectors, it was difficult to distinguish the best at LN detection. VFNet had the highest mAP of 63.91\% over FoveaBox and FCOS, but FCOS had the highest sensitivity of 88.09\% at 4 FP per image. In contrast, the propsed DETR transformer had the highest mAP of 65.41\% and sensitivity of 91.66\% at 4 FP per image. This shows the importance of the self-attention module and spatial positional encoding in the DETR network. We also compared our results against other T2 MRI-based work; compared against Zhao et al. \cite{Zhao2020_mri}, our mAP is 65.41\% vs 64.5\%, and recall is 91.66\% vs 62.6\% at 8 FP. Contrasted with Debats et al. \cite{Debats2019_mri}, our sensitivity is 91.66\% vs 80\% at 8 FP per image. We also compared our results against CT-based work; against Liu et al.\cite{Liu2016_ct}, Seff et al.\cite{Seff2015_ct}, and Roth et al.\cite{Roth2014_ct}, we obtain sensitivities of 91.66\% at 4 FP vs 88\% at 4 FP, 89\% at 6 FP, and 90\% at 6 FP respectively.

We also split our dataset according to the size of the lymph node and evaluated our performance. For LN with SAD $\leq$10mm, the DETR shows a moderate performance of 53.57\% mAP and sensitivity of 87.09\% at 4 FP per image. As shown in Fig. \ref{fig:qual_images}, we attribute the lower mAP to the limitation of the DETR network in detecting small objects \cite{Carion2020_detr}. Our values are lower than Zhao et al\cite{Zhao2020_mri} at $\sim$65\% recall, but we still achieve a much higher recall \textit{without the use of DWI sequences}. But on LN with SAD $\geq$10mm, DETR achieves a mAP of 70.84\% and sensitivity of 94.33\% at 4 FP per image. These results are again consistent with past literature \cite{Zhao2020_mri,Debats2019_mri}, yet we outperform their results with a $\geq$4\% increase in sensitivity. Our proposed DETR executes in 253ms/20s per image/volume vs. 218ms/17s from FCOS, 134ms/10s from FoveaBox, and 276ms/22s from VFNet respectively.

\begin{table}[!h]
\centering\fontsize{9}{12}\selectfont 
\setlength\aboverulesep{0pt}\setlength\belowrulesep{0pt} 
\setlength{\tabcolsep}{7pt} 
\setcellgapes{3pt}\makegapedcells 
\caption{Detection performance of various detectors and our proposed DETR transformer. `S" stands for Sensitivity @[0.5, 1, 2, 4, 6, 8, 16] FP. \mbox{--} indicates unavailable metric values.}
\begin{adjustbox}{max width=\textwidth}
\begin{tabular}{@{} c|c|c|c|c|c|c|c|c @{}} 
\toprule
Method                                                          & mAP       & S@0.5     & S@1       & S@2       & S@4       & S@6       & S@8       & S@16 \\
\midrule

FCOS \cite{Tian2019_fcos}                                       & 60.09     & 61.90     & \textbf{77.38}     & 83.33     & 88.09     & 89.28     & 89.28     & 89.28 \\
FoveaBox \cite{Kong2019_foveabox}                               & 61.67     & 61.90     & 76.19     & 79.76     & 84.52     & 88.09     & 89.28     & 89.28 \\
VFNet \cite{Zhang2021_vfnet}                                    & 63.91     & \textbf{67.85}     & 75        & 80.95     & 83.33     & 83.33     & 83.33     & 83.33 \\

Proposed DETR                            & \textbf{65.41}     & 65.47     & 76.19     & \textbf{88.09}     & \textbf{91.66}    & \textbf{91.66}     & \textbf{91.66}     & \textbf{91.66} \\

\midrule

Proposed DETR (SAD $<$  10mm)                                   & 53.57     & 58.06     & 67.77     & 80.64     & 87.09     & 87.09     & 87.09     & 87.09 \\
Proposed DETR (SAD $\geq$ 10mm)                                 & 70.84     & 69.81     & 79.24     & 92.45     & 94.33     & 94.33     & 94.33     & 94.33 \\

\midrule

Zhao 2020 \cite{Zhao2020_mri} (MRI)                                  & 64.5      & --        & --        & --        & --        & --        & 62.6      & --   \\
Debats 2019 \cite{Debats2019_mri} (MRI)                              & --        & --        & --        & --        & --        & --        & 80        & --   \\
Liu 2016 \cite{Liu2016_ct} (CT)                                    & --        & --        & --        & --        & --        & --        & 88        & --   \\
Seff 2015 \cite{Seff2015_ct} (CT)                                   & --        & --        & --        & --        & --        & 89        & --        & --   \\
Roth 2014 \cite{Roth2014_ct} (CT)                                   & --        & --        & --        & --        & --        & 90        & --        & --   \\

\bottomrule
\end{tabular}
\end{adjustbox}
\label{table_LN_detection_results}
\end{table}

\section{NEW WORK}

We proposed a novel change to the DETR transformer architecture for the challenging task of LN detection in T2 MRI, and merged predictions with weighted boxes fusion to reduce the false positive rate. We also quantified the performance of state-of-the-art detection networks on the LN detection task. Our detection network achieves a clinically acceptable 65.41$\%$ mAP and 91.66$\%$ recall at 4 FP per image. Our results outperform previously published LN detection methods on T2 MRI scans.

\section{CONCLUSIONS}
\label{conclusions}

In this work, we proposed a novel change to the DETR architecture to enable improved LN detection and applied weighted boxes fusion to merge the predicted bounding box clusters. We also quantified the performance of three state-of-the-art one-stage detectors on the LN detection task. Our model was able to detect LN with 65.41$\%$ mAP and 91.66$\%$ recall at 4 FP, and surpassed the prior work in T2 MRI. 

\section{Acknowledgements.} 

This work was supported by the Intramural Research Programs of the NIH Clinical Center and NIH National Library of Medicine. We also thank Jaclyn Burge for the helpful comments and suggestions.

\bibliography{main} 
\bibliographystyle{spiebib} 

\end{document}